\documentclass[conference]{IEEEtran}
\IEEEoverridecommandlockouts
\usepackage{cite}
\usepackage{amsmath,amssymb,amsfonts}
\usepackage{algorithmic}
\usepackage{graphicx}
\usepackage{textcomp}
\usepackage{xcolor}
\usepackage{glossaries}
\newacronym{UAV}{UAV}{Unmanned Aerial Vehicle}
\newacronym{OFDM}{OFDM}{Orthogonal Frequency Division Multiplexing}
\newacronym{OTFS}{OTFS}{Orthogonal Time Frequency Space}
\newacronym{NLOS}{NLOS}{Non-Line-of-Sight}
\newacronym{ECNs}{ECNs}{Emergency Communication Networks}
\newacronym{AONs}{AONs}{Always On Networks}
\newacronym{IoT}{IoT}{Internet of Things}
\newacronym{D2D}{D2D}{Device-to-Device}
\newacronym{WSNs}{WSNs}{Wireless Sensor Networks}
\newacronym{PSNR}{PSNR}{Peak Signal to Noise Ratio}
\newacronym{SNR}{SNR}{Signal-to-Noise Ratio}
\newacronym{QPSK}{QPSK}{Quadrature Phase Shift Keying}
\newacronym{TF}{TF}{Time-Frequency}
\newacronym{SFFT}{SFFT}{Symplectic Finite Fourier Transform}
\newacronym{ISFFT}{ISFFT}{Inverse Symplectic Finite Fourier Transform}
\newacronym{CSI}{CSI}{Channel State Information}
\newacronym{MSE}{MSE}{Mean Squared Error}
\newacronym{KL}{KL}{Kullback-Leibler}
\newacronym{AWGN}{AWGN}{Additive White Gaussian Noise}
\newacronym{ISI}{ISI}{Inter-Symbol Interference}
\newacronym{ICI}{ICI}{Inter-Carrier Interference} 
\newacronym{UE}{UE}{User Equipment} 

\def\BibTeX{{\rm B\kern-.05em{\sc i\kern-.025em b}\kern-.08em
    T\kern-.1667em\lower.7ex\hbox{E}\kern-.125emX}}
\begin{document}

\title{Emergency Communication: OTFS-Based Semantic Transmission with Diffusion Noise Suppression\\
%{\footnotesize \textsuperscript{*}Note: Sub-titles are not captured in Xplore and should not be used}
\thanks{Corresponding authors: Lixin Li, Wensheng Lin.
	
	This work was supported in part by National Natural Science Foundation of China under Grant 62101450, in part by the Young Elite Scientists Sponsorship Program by the China Association for Science and Technology under Grant 2022QNRC001, in part by Aeronautical Science Foundation of China under Grants 2022Z021053001 and 2023Z071053007, in part by the Open Fund of Intelligent Control Laboratory, 
	in part by NSF CNS-2107216, CNS-2128368, CMMI-2222810, ECCS-2302469, US Department of Transportation, Toyota and Amazon.}

\author{
	\IEEEauthorblockN{
		Kexin Zhang\IEEEauthorrefmark{1}, 
		Xin Zhang\IEEEauthorrefmark{1}, 
		Lixin Li\IEEEauthorrefmark{1}, 
		Wensheng Lin\IEEEauthorrefmark{1}, 
		Wenchi Cheng\IEEEauthorrefmark{2} 
		and Qinghe Du\IEEEauthorrefmark{3}} 
		\IEEEauthorblockA{\IEEEauthorrefmark{1}School of Electronics and Information, Northwestern Polytechnical University, Xi’an, China, 710129}
		\IEEEauthorblockA{\IEEEauthorrefmark{2}State Key Laboratory of Integrated Services Networks, Xidian University, Xi’an, China, 710071}
		\IEEEauthorblockA{\IEEEauthorrefmark{3}School of Information and Communications Engineering, Xi’an Jiaotong University, Xi’an, China, 710049} 
	} 
} 
\maketitle

\begin{abstract}
Due to their flexibility and dynamic coverage capabilities, \glspl{UAV} have emerged as vital platforms for emergency communication in disaster-stricken areas. However, the complex channel conditions in high-speed mobile scenarios significantly impact the reliability and efficiency of traditional communication systems. This paper presents an intelligent emergency communication framework that integrates \gls{OTFS} modulation, semantic communication, and a diffusion-based denoising module to address these challenges. \gls{OTFS} ensures robust communication under dynamic channel conditions due to its superior anti-fading characteristics and adaptability to rapidly changing environments. Semantic communication further enhances transmission efficiency by focusing on key information extraction and reducing data redundancy. Moreover, a diffusion-based channel denoising module is proposed to leverage the gradual noise reduction process and statistical noise modeling, optimizing the accuracy of semantic information recovery. Experimental results demonstrate that the proposed solution significantly improves link stability and transmission performance in high-mobility UAV scenarios, achieving at least a 3dB SNR gain over existing methods.
\end{abstract}

\begin{IEEEkeywords}
Semantic communication, diffusion model, OTFS
\end{IEEEkeywords}

\section{Introduction}
In recent years, the world has faced numerous devastating natural and artificial disasters, such as earthquakes, hurricanes, and tsunamis, resulting in significant loss of life and property. In the first half of 2024 alone, global natural disasters caused an estimated \$120 billion in damages, exceeding the decade's average. Disasters often cripple communication infrastructure, disrupting rescue operations and leaving traditional systems inoperative. Thus, establishing reliable emergency communication networks is crucial to facilitate efficient disaster response\cite{sec1-1}. 

High-mobility platforms, such as Unmanned Aerial Vehicle (UAV)-based emergency communication systems\cite{sec1-2}, offer unique advantages due to their flexibility, rapid deployment, and ability to provide real-time, reliable data transmission and collaborative communication across disaster-stricken areas\cite{sec1-3}. However, these advantages come with significant challenges, particularly in scenarios involving high-speed mobility and dynamic environments, where traditional communication techniques often fall short.

Specifically, traditional communication methods like \gls{OFDM} face limitations in time-varying channels, struggling with signal attenuation and performance degradation under complex multipath conditions. To address these challenges, Orthogonal Time Frequency Space (OTFS) modulation has emerged as a promising solution. OTFS effectively counters the detrimental effects of Doppler shifts and multipath fading, transforming complex, time-varying channels into sparse and more manageable representations by operating in the time-delay Doppler domain. This ensures stable and reliable communication, even in \gls{NLOS} and high-mobility scenarios.

Building on these advancements, integrating semantic communication \cite{sc} into OTFS-based UAV networks presents a further step toward overcoming the limitations of existing communication frameworks in disaster relief operations. By prioritizing task-relevant content, semantic communication reduces data redundancy and enhances decision-making efficiency in critical scenarios. Accordingly, this research contributes to the field in the following ways:

\begin{itemize}
	\item This study pioneers the integration of OTFS modulation with semantic communication frameworks, leveraging OTFS’s robustness against channel fading in dynamic environments. The proposed method establishes a fusion model of semantic transmission and physical-layer modulation, offering a practical and effective solution for high-mobility and dynamic scenarios, such as vehicular networks and UAV-based communication systems.

	\item A novel channel denoising framework combines the iterative noise reduction capability of diffusion processes with the statistical characteristics of channel noise. This approach accurately generates and matches the probability distribution of channel noise, significantly improving denoising performance during transmission and achieving dual optimization in noise elimination and semantic recovery.
	
	\item Experiments validate the proposed framework, demonstrating its superiority in transmission efficiency, link stability, and task-specific performance in OTFS-based semantic communication systems, highlighting the integrated framework's potential for highly dynamic communication environments.
\end{itemize}

\begin{figure*}[!t]
	\centering
	\includegraphics[width=\textwidth]{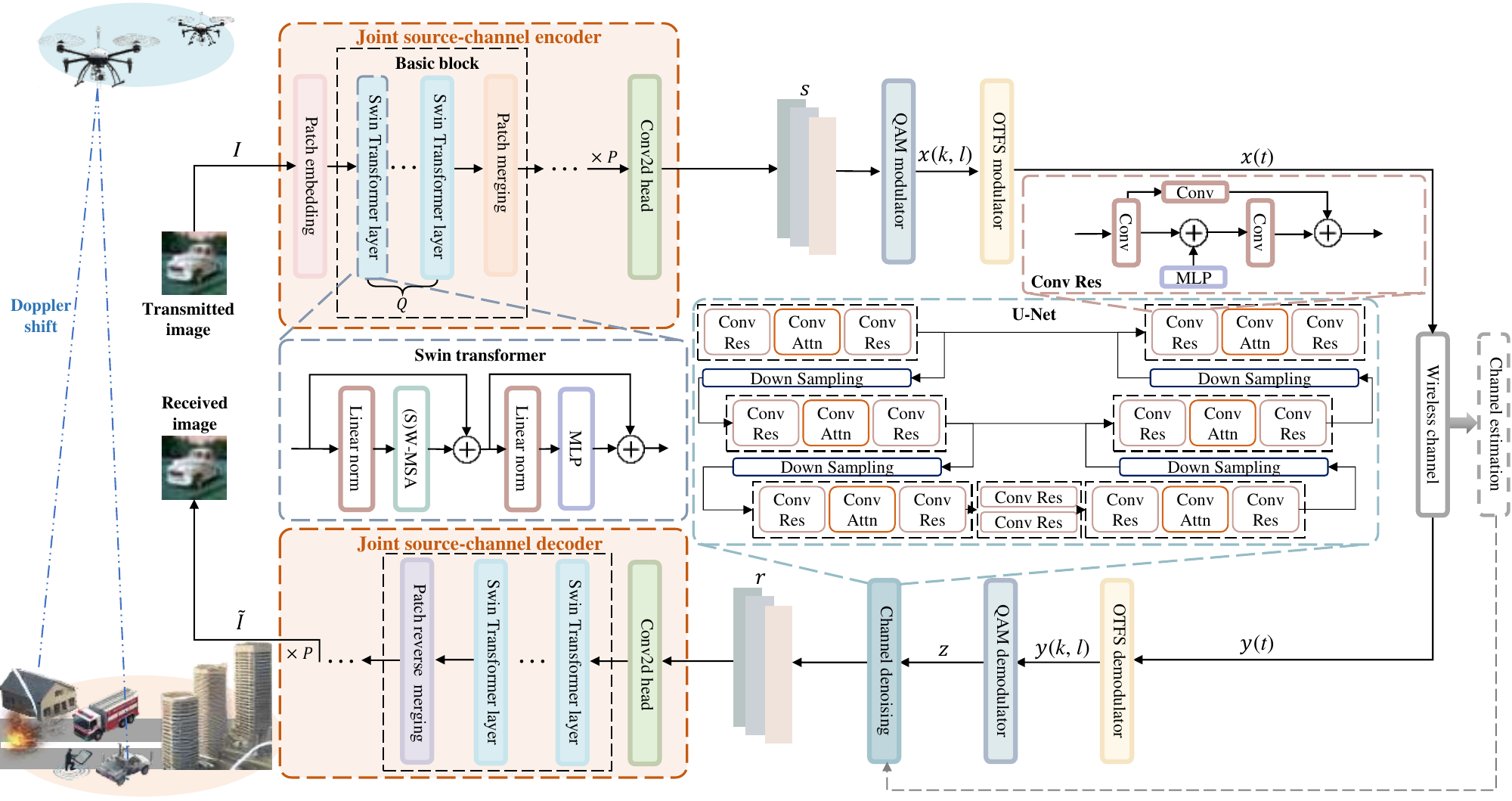}
	%\textcolor{blue}{\caption{The overview of the proposed generative AI semantic communication system.}}
	\caption{The architecture of the proposed method.}
	\label{fig1}
\end{figure*}

\section{Related Work}

In the aftermath of natural or human-made disasters, traditional communication infrastructures frequently become overwhelmed, leading to partial or complete outages. This reality emphasizes the pressing need for reliable emergency networks that ensure continuous coverage and maintain communication during critical moments. In recent years, a wave of research has emerged to confront these urgent challenges head-on. Governments around the globe recognize this necessity and are investing heavily in developing robust emergency communication systems and advancing related technologies, all intending to enhance disaster response.

Research on \gls{ECNs} has focused on various technological advancements. Wang \textit{et al.}\cite{38} reviewed ECNs and big data analytics, highlighting the importance of spatial and content-based analysis to optimize resource allocation in disaster areas. Pervez \textit{et al.}\cite{39} explored wireless emergency response systems, comparing different wireless technologies and their applications in disaster scenarios. Ali \textit{et al.} \cite{40} discussed emerging technologies such as \gls{IoT}, \gls{D2D} communication, \glspl{UAV}, and \gls{WSNs}, emphasizing their potential in disaster management. Du \textit{et al.}\cite{41} provided a systematic review of emergency management networks, focusing on research trends and methodologies. Debnath \textit{et al.}\cite{42} examined the performance challenges of emergency communication networks, particularly energy efficiency and connectivity. Anjum \textit{et al.}\cite{43} categorized communication strategies in mobile ad hoc networks, focusing on routing protocols for disaster situations. Sharma \textit{et al.}\cite{44} investigated UAV-based communication technologies, including machine learning and path planning for enhanced drone communication. Kishorbhai \textit{et al.}\cite{45} analyzed \gls{AONs} and their potential to provide reliable communication during emergencies.

While these studies offer valuable contributions, they primarily focus on the development and performance optimization of communication infrastructures. However, current research often lacks a focus on the semantic and contextual relevance of transmitted information during emergencies. Beyond ensuring data integrity, the efficiency and relevance of transmitted information are crucial for supporting real-time decision-making. Addressing this gap, recent advancements are exploring semantic communication, which prioritizes the meaning of data rather than its raw transmission. Building on this, by integrating semantic communication with AI, critical information, such as the locations of casualties or fire sources, can be identified, enhanced, and transmitted efficiently. This emerging paradigm reduces communication overhead and enhances operational efficiency in rescue and disaster management efforts, such as supply drops and personnel rescues.

\section{System Model}
The proposed intelligent emergency communication system comprises a semantic coding module, a physical layer modulation module, and a channel denoising module, as illustrated in Fig.~\ref{fig1}.

\subsection{Semantic Coding Module}\label{AA}
Let $I$ denote the source image, and $\widetilde{I}$ represent the reconstructed RGB images. 
The semantic encoder is designed with a hierarchical architecture based on the Swin Transformer backbone\cite{sec3-witt}, whose specific structure is shown in the orange box in Fig.~\ref{fig1}. 
The encoder employs a patch embedding module to divide the input image into patches, followed by progressively deeper Swin Transformer blocks for hierarchical feature extraction. 
Each basic block employs shifted window multi-head self-attention to capture the image's local and global dependencies. 
An adaptive modulator is introduced at specific stages to refine features, and the output layer maps the features into a $C$-dimensional latent space using a $1 \times 1$ convolution layer. 
The decoder reverses these operations, employing Swin Transformer layers with upsampling and patch merging to restore the spatial dimensions and details of the input image. 
This symmetric design ensures that the decoder efficiently utilizes the latent representation generated by the encoder for accurate reconstruction, leveraging key innovations such as adaptive modulation and hierarchical attention mechanisms.
 
The semantic feature vectors can be expressed as:
\begin{equation}
	s=\boldsymbol{E}\left(I ; \varphi_\alpha\right) \in \mathbb{R}^{h \times w \times c},
\end{equation}
where $\boldsymbol{E}\left(\cdot \right)$ denotes the semantic encoder network, $\varphi_\alpha$ is the parameter set of the corresponding encoder, $h$ and $w$ represent the height and width of the input image, respectively, and $c$ is the dimension of the input images.
%and $n$ is the dimensionality of codes with a value of 256.

The semantic decoder processes the received signal disturbed during transmission, described as:
\begin{equation}
	\widetilde{I}=\boldsymbol{E}^{-1}\left(r; \varphi_\gamma\right) \in \mathbb{R}^{h \times w \times c},
\end{equation}

\noindent where $\boldsymbol{E}^{-1}(\cdot)$ denotes the semantic decoder network, with $\varphi_\gamma$ as its parameter set.

The quantization process is essential in practical systems where media is transmitted using symbols from a finite alphabet. Although often omitted in descriptions for simplicity, quantization is implicitly applied when converting image pixels or feature values from continuous analog signals to fixed-length binary sequences suitable for transmission.

\subsection{Physical Layer Modulation}
\paragraph{Definition}
In OTFS modulation, the basis for signal representation is characterized by delay \(\tau\) and Doppler shift \(\nu\). These parameters are inversely related to the subcarrier spacing \(\Delta f\) and the symbol duration \(T\).

To map signals between the time-frequency domain and the delay-Doppler domain, OTFS employs two essential transformations: the \gls{SFFT} transforms the signal from the time-frequency domain to the delay-Doppler domain. In contrast, the \gls{ISFFT} performs the reverse operation.

\paragraph{Modulation}
The OTFS modulation process starts by mapping binary bits to QAM-modulated symbols. These symbols are arranged in a delay-Doppler data grid $x[k, l]$, and the ISFFT is applied to transform them into time-frequency domain symbols \(X[n, m]\):
\begin{equation}
	X[n, m] = \text{ISFFT}\left(x[k, l]\right).
\end{equation}

Subsequently, a dynamic windowing mechanism is employed in the time-frequency domain to enhance signal robustness and mitigate interference. Handling data frames and normalization in the implementation indirectly represents this windowing effect. Specifically, the data grid is padded and adjusted during the modulation process to ensure the proper mapping of symbols within the time-frequency plane.

The generated \(X[n, m]\) is mapped to the time domain using the Heisenberg transform, producing the transmitted signal \(x(t)\):
\begin{equation}
	x(t) = \sum_{n=0}^{N-1} \sum_{m=0}^{M-1} X[n, m] \varphi_{\text{tx}}(t - nT) e^{j 2\pi m \Delta f (t - nT)},
\end{equation}
where \(\varphi_{\text{tx}}(t)\) is the transmit pulse shaping function. 

\paragraph{Channel Propagation}
The transmitted signal \(x(t)\) propagates through a delay-Doppler domain channel, with the received signal expressed as:
\begin{equation}
	y(t) = \int_\nu \int_\tau h(\tau, v) x(t-\tau) e^{j 2\pi v(t-\tau)} \, \mathrm{d}\tau \, \mathrm{d}v,
\end{equation}
where \(h(\tau, v)\) represents the channel's delay and Doppler response. The channel captures multipath propagation and time-varying Doppler effects.

\paragraph{Demodulation}
At the receiver, the received time-domain signal \(y(t)\) is transformed into the time-frequency domain through TF demodulation, yielding \(Y[n, m]\). The signal is then processed with the receive window \(W_{\text{r}}[n, m]\), which reduces interference and ensures alignment in the delay-Doppler domain. Specifically, signal merging from multiple paths is performed using weighted operations, such as Maximum Ratio Combining (MRC) and noise suppression techniques. These processes effectively mimic the function of the receive window by aligning the signals and mitigating interference.

The \gls{SFFT} is applied to the processed signal to obtain the demodulated delay-Doppler symbols \(y[k, l]\):
\begin{equation}
	y[k, l] = \text{SFFT}\left(Y_p[n, m]\right), \quad k = 0, \ldots, N-1, \, l = 0, \ldots, M-1.
\end{equation}

The relationship between the transmitted and received symbols is given by:
\begin{equation}
	y[k, l]= \frac{1}{M N} \sum_{m=0}^{M-1} \sum_{n=0}^{N-1} x[n, m] \cdot h_w\left(\frac{k-n}{N T}, \frac{l-m}{M \Delta f}\right) + v[k, l],
\end{equation}
where \(v[k, l]\) represents the \gls{AWGN}, and \(h_w\) is the effective channel response.

Finally, the demodulated symbols \(y[k, l]\) are mapped to binary bits through QPSK demodulation.

\subsection{Channel Denoising Module}
The channel denoising module complements traditional channel equalization by further removing residual noise, enhancing the communication system's overall performance. It consists of two key processes: the forward diffusion process and the reverse inference process.
\paragraph{Forward Diffusion Process Design}
As proven in \cite{sec3-ccdm}, the forward diffusion process resembles channel equalization and normalization operations in wireless communications, providing a theoretical basis for its application in channel noise mitigation. This process progressively adds noise to the input signal \(z_0\) over \(T\) steps, transforming it into \(z_t\), defined as:
\begin{equation}
	z_t = \sqrt{\alpha_t} z_{t-1} + \sqrt{1 - \alpha_t} W_n \epsilon,
\end{equation}
where \(\epsilon \sim \mathcal{N}(0, I)\) is Gaussian noise, \(W_n\) represents the noise coefficient matrix capturing channel characteristics, and \(\alpha_t\) controls the noise-signal ratio. To simplify computation, the process can be reparameterized as:
\begin{equation}
	z_t = \sqrt{\bar{\alpha}_t} z_0 + \sqrt{1 - \bar{\alpha}_t} W_n \epsilon,
\end{equation}
where \(\bar{\alpha}_t = \prod_{i=1}^t \alpha_i\) is the cumulative product of the noise schedule parameters.

\paragraph{Reverse Inference Process Design} 
The reverse inference process is to iteratively reconstruct the original signal \( z_0 \) from the noise-corrupted signal \( y_r \) by leveraging the learned noise characteristics. At each step \(t\), the model predicts the noise component \(\epsilon_\theta(z_t, h_r, t)\), where \(h_r\) represents \gls{CSI}, and refines \(z_{t-1}\). The process begins with \(z_m = y_r\) and iteratively reconstructs \(z_0\) over \(m\) steps. Each step is modeled as follows:
\begin{equation}
	z_{t-1} = \sqrt{\alpha_{t-1}} z_0 + \sqrt{1 - \alpha_{t-1}} W_n \epsilon_\theta(z_t, h_r, t).
\end{equation}

The number of steps \(m\) is optimized by minimizing the \gls{KL} divergence to balance computational efficiency and performance. In the final step (\(t = 1\)), the reconstructed signal is:
\begin{equation}
	r = \frac{1}{\sqrt{\alpha_1}} \left( z_1 - \sqrt{1 - \alpha_1} W_n \epsilon_\theta(z_1, h_r, 1) \right),
\end{equation}
here, \( r \) representing the clean, denoised signal.

By leveraging learned noise characteristics and channel properties, the reverse inference process ensures accurate reconstruction of the transmitted signal while effectively mitigating noise.

\subsection{Training Objective}
The proposed system framework is trained in three stages to ensure optimal performance. In the first stage, we focus solely on training the semantic communication module, including the semantic encoder and decoder. The semantic encoder extracts key features from the input data and compresses them into a compact representation, then decoded by the semantic decoder.

The training objective is to minimize the reconstruction error between the original input \(I\) and the decoded output \(\hat{I}\), using the mean squared error (\gls{MSE}) as the primary loss function. Additionally, a \gls{KL} divergence penalty is introduced to regularize the encoder’s output, encouraging statistical alignment with a standard normal distribution. The loss function for this stage is defined as:
\begin{equation}
	\mathcal{L}_1(\phi, \psi) = \mathbb{E}_{I \sim p_I, \hat{I} \sim p_{\hat{I}}} \left[ \| I - \hat{I} \|^2_2 \right],
\end{equation}
where \(\phi\) and \(\psi\) represent the semantic encoder and decoder parameters, respectively. 

The diffusion-based channel denoising module, integrated into joint end-to-end training in the second stage, aims to enhance the system's anti-interference capability by simulating channel impairments. Here, the parameters of the encoder are fixed, allowing the denoising module to focus on learning the distribution of \(x_0\). The training objective is to minimize the difference between the actual noise \(\epsilon\) and the predicted noise \(\epsilon_\theta(z_t, h_r, t)\), where \(\theta\) represents the model parameters and \(h_r\) denotes the normalized channel state information. The corresponding loss function is given by:
\begin{equation}
	\mathcal{L}_2(\theta) = \mathbb{E}_{z_0, \epsilon, t} \| \epsilon - \epsilon_\theta(z_t, h_r, t) \|^2.
\end{equation}

Finally, the semantic decoder is re-trained jointly with the trained semantic encoder and denoising module to minimize \(d(I; \widetilde{I})\), which measures the semantic similarity between the original input \(I\) and the reconstructed output \(\widetilde{I}\). During this stage, the entire joint system is performed through the actual channel, while only the decoder parameters are updated.

\section{Demonstration And Results}
In this section, we present the experimental setup to validate the effectiveness of the proposed intelligent emergency communication framework. Specifically, the CIFAR-10 dataset is used, and the Adam optimizer is employed with a learning rate of \(1 \times 10^{-4}\). Training in the semantic network is conducted with a batch size of 32 under different channel conditions with a fixed \gls{SNR} of 13 dB. The channel denoising module is implemented based on a U-Net architecture, as described in \cite{sec4-unet}, with inputs \( z \) (noise-corrupted signal) and \( h_r \) (channel state information). The forward diffusion process is set to \( T = 200 \), with noise scheduling parameters \(\alpha_t\) linearly decreasing from \(\alpha_1 = 0.9999\) to \(\alpha_T = 0.98\). 
For the OTFS modulation, the parameters \(N\) and \(M\), representing the number of subcarriers and symbols per frame, are configured as \(N = 128\) and \(M = 256\). The testing is performed across SNRs ranging from 0 to 15 dB to evaluate the robustness of the proposed framework under varying noise levels.

We adopt the the classical separation-based scheme as a benchmark for performance comparison, and JPEG uses LDPC for channel coding, and the modulation order is set to 4.

\begin{figure}[htbp]
	\centerline{\includegraphics[width=3in]{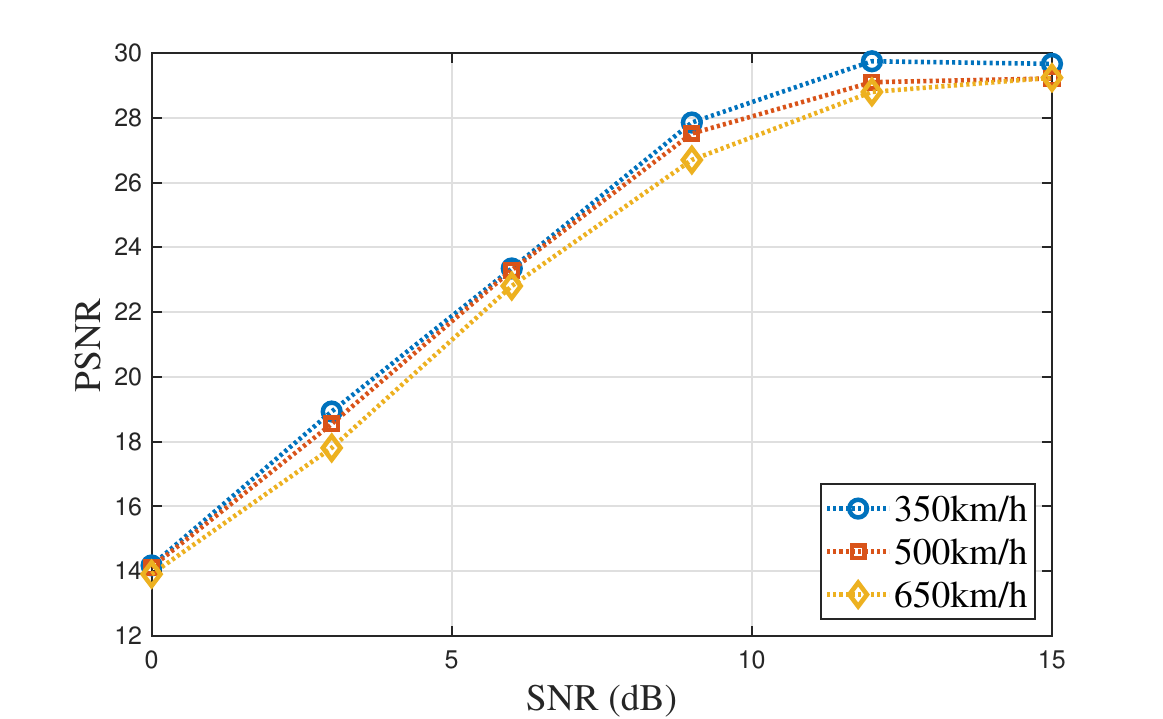}}
	\caption{Performance comparison across SNR levels at UE speeds of 350 km/h, 500 km/h, and 650 km/h (SCS = 15 kHz).}
	\label{fig2}
\end{figure}

In this study, we evaluated the performance of the proposed system under different SNR conditions and varying \gls{UE} speeds. Fig.~\ref{fig2} demonstrates that the system achieves consistent improvements in PSNR across all SNR levels, regardless of speed. At 350 km/h, the PSNR curve exhibits the best performance, with a relatively smooth increase as the SNR improves. Although the performance slightly declines due to increased Doppler effects at 500 km/h and 650 km/h, the system still maintains a high PSNR, particularly at low SNRs. For instance, at 6 dB, the system achieves a PSNR above acceptable thresholds even at higher speeds. This indicates that the integration of OTFS modulation and semantic denoising enhances the system's ability to mitigate channel impairments and provides resilience against velocity-induced Doppler spread.

\begin{figure}[htbp]
	\centerline{\includegraphics[width=3in]{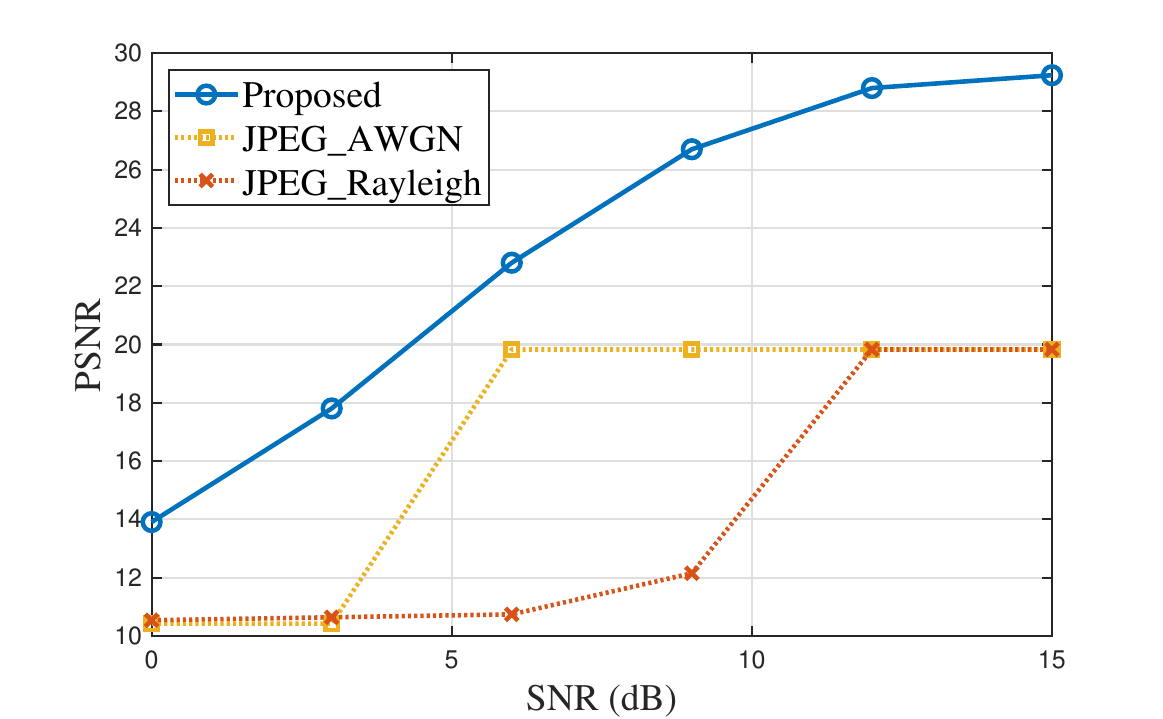}}
	\caption{PSNR performance across different channels and methods.}
	\label{fig3}
\end{figure}

As depicted in Fig.~\ref{fig3}, the proposed method is evaluated under the inherently more demanding OTFS channel, unlike JPEG, which is tested on simpler AWGN and Rayleigh channels. Despite significantly harsher conditions, the proposed system consistently demonstrates exceptional robustness and adaptability, particularly in high-mobility scenarios with user velocities reaching 650 km/h. The resilience of the proposed method can be attributed to two key factors: the OTFS modulation, which efficiently mitigates high Doppler shifts and severe time-frequency dispersion, and the diffusion model, which effectively learns and denoises the complex channel noise characteristics, preserving communication reliability.

\section{Summary}
This work presents a novel emergency communication framework to address the challenges of high-mobility disaster scenarios. Firstly, semantic communication improves efficiency by reducing redundant data. Next, OTFS modulation is employed at the transmitter to transform time-varying channels into a delay-Doppler domain, ensuring stable signal transmission by mitigating the effects of Doppler shifts and multipath fading. The received signal then passes through a diffusion-based denoising module, which gradually refines the signal to recover the transmitted semantic information accurately. This multi-stage approach ensures that critical information is efficiently transmitted and reliably recovered in dynamic and challenging environments, thereby addressing the limitations of conventional communication systems in disaster relief operations.

\end{document}